\documentclass[sigconf,screen]{acmart}

\usepackage[utf8]{inputenc}
\usepackage[T1]{fontenc}
\usepackage{amsmath,amsfonts,nicefrac,,gensymb}
\usepackage{hyperref,url,booktabs,multirow,microtype}
\usepackage{enumitem,algorithm,algorithmicx,algpseudocode}
\usepackage{stfloats,xcolor,graphicx}
\usepackage{soul}

\copyrightyear{2022}
\acmYear{2022}
\setcopyright{rightsretained}
\acmConference[ETRA '22]{2022 Symposium on Eye Tracking Research and Applications}{June 8--11, 2022}{Seattle, WA, USA}
\acmBooktitle{2022 Symposium on Eye Tracking Research and Applications (ETRA '22), June 8--11, 2022, Seattle, WA, USA}
\acmDOI{10.1145/3517031.3531630}
\acmISBN{978-1-4503-9252-5/22/06}


\begin{document}

\title{Multidisciplinary Reading Patterns of Digital Documents}

\author{Bhanuka Mahanama}
\email{bhanuka@cs.odu.edu}
% \affiliation{\department{Department of Computer Science}\institution{Old Dominion University}\city{Norfolk}\state{VA}\country{USA}}

\author{Gavindya Jayawardena}
\email{gavindya@cs.odu.edu}
\affiliation{\department{Department of Computer Science}\institution{Old Dominion University}\city{Norfolk}\state{VA}\country{USA}}

\author{Yasasi Abeysinghe}
\email{yasasi@cs.odu.edu}
% \affiliation{\department{Department of Computer Science}\institution{Old Dominion University}\city{Norfolk}\state{VA}\country{USA}}

\author{Vikas Ashok}
\email{vganjigu@odu.edu}
\affiliation{\department{Department of Computer Science}\institution{Old Dominion University}\city{Norfolk}\state{VA}\country{USA}}

\author{Sampath Jayarathna}
\email{sampath@cs.odu.edu}
\affiliation{\department{Department of Computer Science}\institution{Old Dominion University}\city{Norfolk}\state{VA}\country{USA}}

\begin{abstract}
Reading plays a vital role in updating the researchers on recent developments in the field, including but not limited to solutions to various problems and collaborative studies between disciplines. Prior studies identify reading patterns to vary depending on the level of expertise of the researcher on the content of the document.
% New
We present a pilot study of eye-tracking measures during a reading task with participants with different domain expertise 
to characterize their reading patterns. 
% Old
% We present a pilot study of eye-tracking measures during a reading task with participants across different areas of expertise with the intention of characterizing the reading patterns using both eye movement and pupillary information. 
\end{abstract}


\maketitle

\section{Introduction}

The reading patterns of 
digital documents
(scholarly articles)
vary from person to person across various disciplines. Despite the consensus that reading patterns are stochastic, recent studies identify similarities between individuals with common expertise. The studies by \cite{jayawardena2020reading-patterns, mahanama2021-reading-patterns-jcdl} identify that participants spend the most time in the methodology section, with a relatively low cognitive load.  

However, these studies only rely on pilot studies of participants from the computer science domain. As a result, the findings of the studies can be questionable for other disciplines. 
Therefore, we present a dataset \footnote{\url{https://github.com/nirdslab/Multidisciplinary-Reading-Patterns}}
that includes eye-tracking behaviors of researchers from multiple disciplines.
Our contributions are, 
% The contributions of our study are as follows,
\begin{enumerate}
    % \item Present a dataset that includes eye-tracking behaviors of
    % novice to expert researchers.
    % researchers from multiple disciplines.
    \item Conduct a preliminary analysis on the generalizability of claims of previous studies across domains.
    \item Discuss implications of our preliminary results and potential research avenues.
    % \item Discuss the potential usage of our dataset and findings towards digital documents and applications.
\end{enumerate}

\section{Methodology}

% \noindent
% \textbf{Data Collection : }
We recruited seven (6 F, 1 M)
% volunteer 
graduate students as researchers in Computer Science (CS) (2), Mathematics (2), and Physics (3). 
The participants aged between 25 and 35 years, with research experience ranging from one to five years. We verbally confirmed their experience in reading research papers and verified their vision through a visual acuity test. 
% Old
% The research experiences of the participants ranged between one year to five years and aged between 25 - 35 years. 
% We confirmed their experience in reading research papers in conference venues verbally. 
% We verified the vision of the participants through a visual acuity test. 

\begin{figure}[t]
\centering
    \includegraphics[width=.33\linewidth]{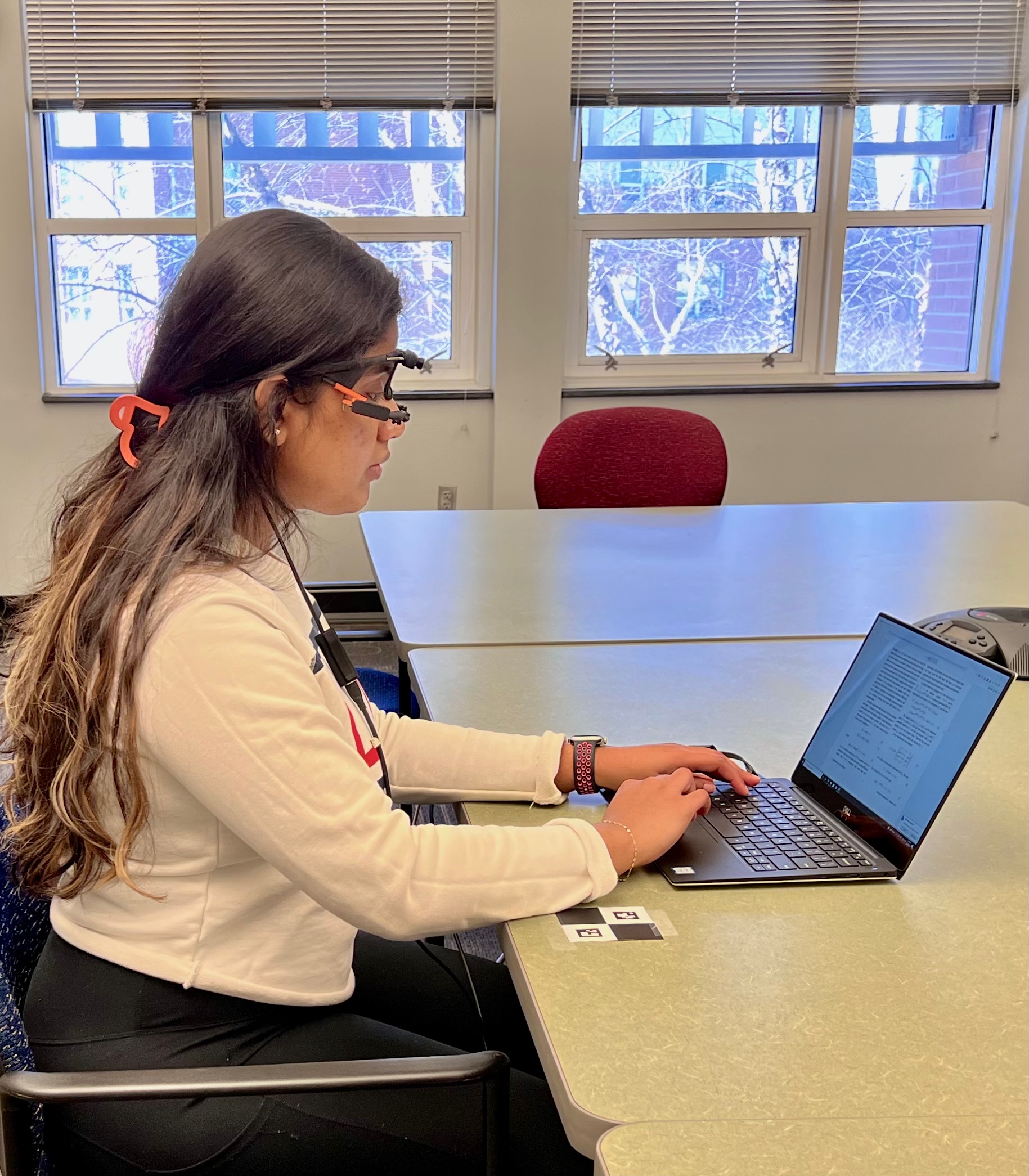}
    \hfill
    \includegraphics[width=.6\linewidth]{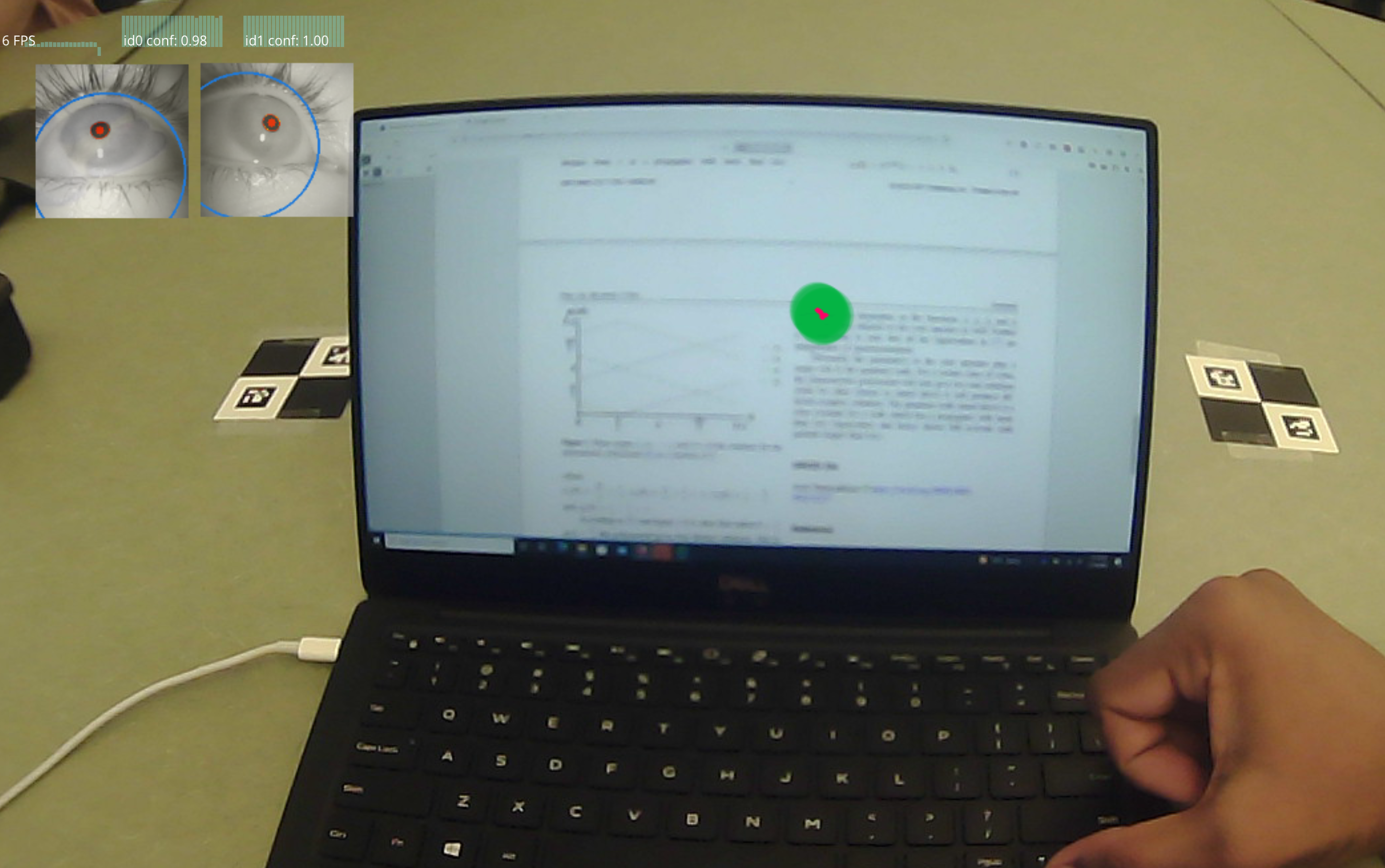}
    \caption{\textbf{Experimental setup.} Left: Participant during reading task, Right: Sample recording.
    % (Circle indicate gaze position).
    \label{experiment-setup}}
\end{figure}

% \noindent
% \textbf{Experimental Setup : }
We selected two articles of two pages in Computer Science and Physics for the reading task. 
After reading each paper, 
% we asked each participant to briefly summarize the article verbally and answer queries by the proctor to confirm their understanding. 
each participant briefly summarized the article verbally and answered queries by the proctor to confirm their understanding.
We allowed the participants to perform the task in a laboratory setting with their preferred lighting, brightness, and zooming levels (See Figure \ref{experiment-setup} for the experimental setup). On a given day, we limited the experiment to a single paper per participant to eliminate the effects of fatigue in the dataset. 

We used PupilLabs Core \footnote{\url{https://pupil-labs.com/products/core/}} eye-tracker to record eye-movements at a sampling frequency of 120 Hz with an accuracy of $0.60 \degree$.
Each participant was calibrated using the 5-point calibration in Pupil Capture
\footnote{\url{https://docs.pupil-labs.com/core/software/pupil-capture/}} 
and confirmed the accuracy through manual tests. 
For each paper, we annotated five sections: (1) title, (2) abstract, (3) introduction and related work, (4) methodology, and (5) figures. 
We contacted the author and confirmed the section classification for the articles without explicitly defined sections.

% \begin{figure}[t]
% \centering
% \includegraphics[width=0.6\linewidth]{images/experiment-setup.jpeg}
% \caption{Experimental setup\label{experiment-setup}}
% \end{figure}

% \begin{figure*}[t]
% \centering
%     \includegraphics[width=.25\linewidth]{images/experiment-setup.jpeg}
%     \includegraphics[width=.46\linewidth]{images/sample-recording.png}
%     \caption{\textbf{Experimental setup.} Left: Participant during reading task, Right: Sample recording (Circle illustrate gaze position).\label{experiment-setup}}
% \end{figure*}

% \noindent
% \textbf{Data Analysis : }
For each user session, we replayed the gaze positions using Pupil Player
% \footnote{https://docs.pupil-labs.com/core/software/pupil-player/}
and annotated the eye movements in each of the sections 
% identified above 
% for the section of the paper 
with a forum of three manual annotators. 
% \todo{Select relevant metrics}
Then we extracted the annotated data from Pupil Player software and utilized $x,y,timestamp,$ and $pupil$ $diameter$ within each section to generate multiple eye gaze metrics.
Based on the annotations, we computed 
% different metrics including, 
fixation count, 
fixation duration, 
% saccade amplitude,
Low/High Index of Pupillary Activity (LHIPA) \cite{duchowski2020LHIPA}, 
and
average pupil diameter.
% and percentage change of pupil diameter. 
We calculated the aforementioned eye gaze metrics for each participant, for each paper, for each section including revisits to the same section.
% When calculating eye gaze metrics related to pupil dilation, we took the visiting order into account in order to preserve the temporal information.

\section{Results}

\begin{figure}[t]
\centering
    \fbox{
        \includegraphics[width=.44\linewidth]{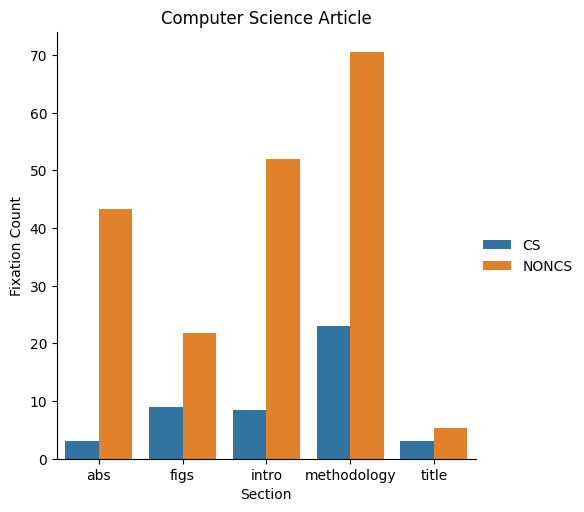}
        % \caption{a}
    }
    \hfill
  \fbox{
        \includegraphics[width=.44\linewidth]{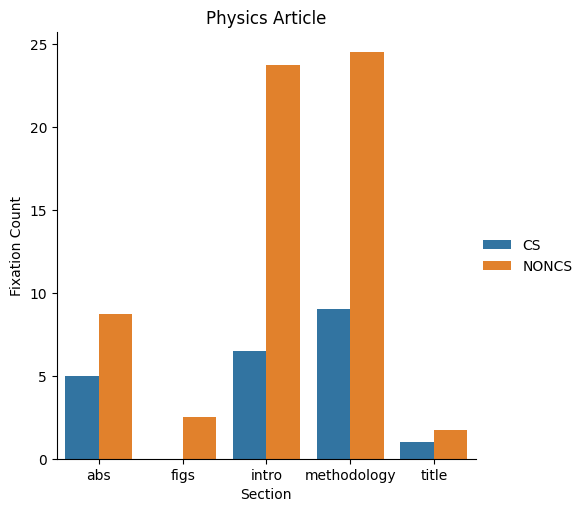}
        % \caption{b}
    }
    \caption{\textbf{Average Fixation Count} \label{fixation-count}}
\end{figure}

\begin{figure}[t]
\centering
    \fbox{
        \includegraphics[width=.44\linewidth]{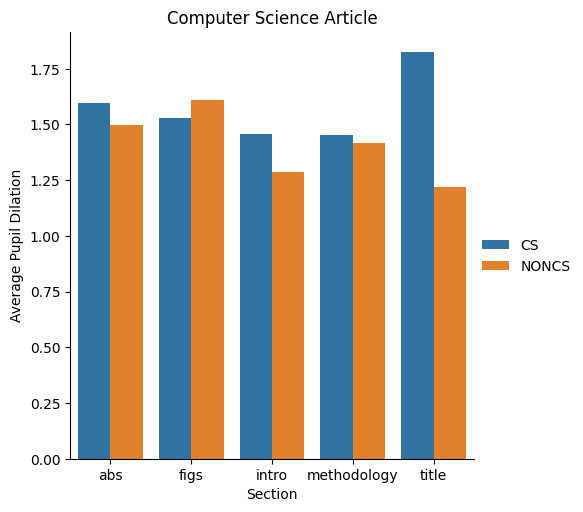}
        % \caption{a}
    }
    \hfill
  \fbox{
        \includegraphics[width=.44\linewidth]{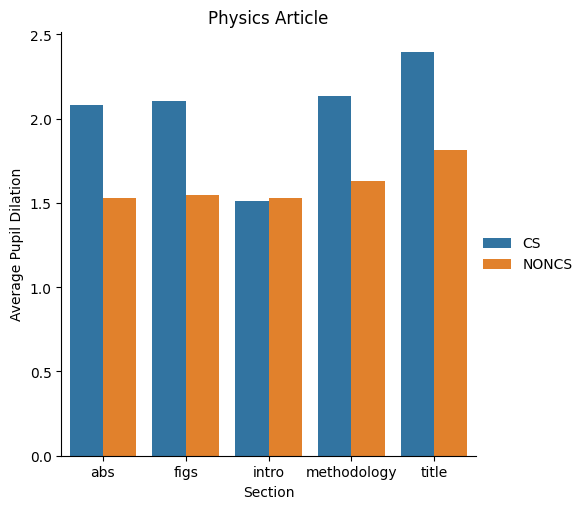}
        % \caption{b}
    }
    \caption{\textbf{Average Pupil Dilation} \label{pupil-dilation}}
\end{figure}

\begin{figure}[t]
\centering
    \fbox{
        \includegraphics[width=.44\linewidth]{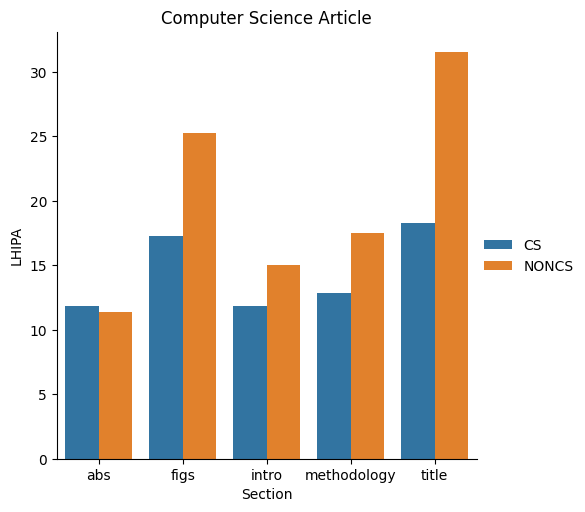}
        % \caption{a}
    }
    \hfill
  \fbox{
        \includegraphics[width=.44\linewidth]{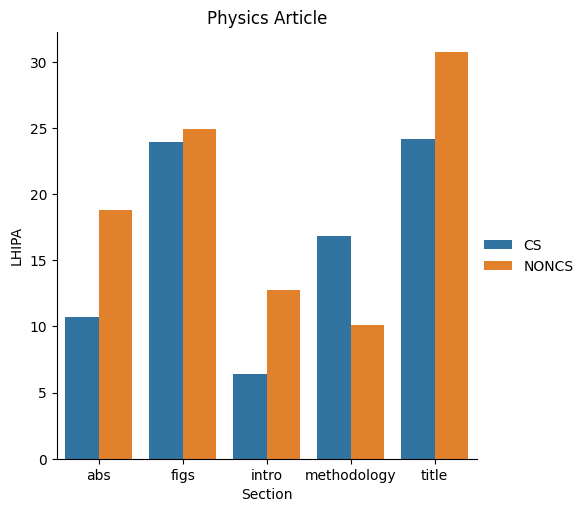}
        % \caption{b}
    }
    \caption{\textbf{Low/High Index of Pupillary Activity (LHIPA)} \label{lhipa}}
\end{figure}

% \begin{figure}[t]
% \centering
%     \fbox{
%         \includegraphics[width=0.8\linewidth]{figures/combined-measures.png}
%         % \caption{a}
%     }
%     \caption{Top: Average Fixation Count, Center: Average Pupil Dilation, Bottom:  Low/High Index of Pupillary Activity (LHIPA) \label{lhipa}}
% \end{figure}

% \begin{figure}[t]
% \centering
%     \fbox{
%         \includegraphics[width=0.75\linewidth]{figures/combined-measures-updated.png}
%         % \caption{a}
%     }
%     \caption{Top: Average Fixation Count, Center: Average Pupil Dilation, Bottom: LHIPA. \label{lhipa}}
% \end{figure}

% \noindent
% \textbf{Fixation Count: }
Our preliminary results indicate the participants fixated more on the methodology sections of the paper irrespective of the domain consistent with the past studies\cite{jayawardena2020reading-patterns, mahanama2021-reading-patterns-jcdl}. 
Considering the computer science article, Non-CS participants have more fixation counts across all sections than CS participants. 
We suspect familiarity with the content as a possible reason for the behavior. 
Moreover, we observe similar behavior (fixation count ranking) among Non-CS participants while reading the physics article, with much lesser fixations. 

Following the behavior of Non-CS participants, we expected CS participants to exhibit a higher number of fixations while reading the physics article. 
On the contrary, we observed the number of fixations decrease while being lesser than Non-CS, indicating less time spent. 
We presume the CS participants have only skimmed through the content combined with the unfamiliar content. 

% OLD Content
% 
% Our preliminary results indicate the participants fixated more on the methodology sections of the paper irrespective of the domain consistent with the past studies \cite{jayawardena2020reading-patterns, mahanama2021-reading-patterns-jcdl}. 
% When reading the computer science article, Non-CS participants have fixated more on the abstract, introduction, and methodology sections than CS participants, indicating unfamiliarity with the content causing the behavior. 
% On the other hand, CS participants, though they have have the highest number of fixations on methodology section, compared to the Non-CS participants, they have not fixated (spent too much time) on the 
% computer science article. This behavior yields the familiarity of the field. 

% In physics article, Non-CS participants have fixated the most in abstract, introduction, and methodology sections. But, compared to computer science article, 
% Non-CS participants fixated on those sections much less while reading the physics article.

% Similar to the fixation count behavior observed while Non-CS participants reading the computer science article, we expected CS participants to have higher fixation counts compared to Non-CS participants while reading the physics article. 
% On the contrary, we observed that CS participants fixated much less on all of the sections of physics article compared to Non-CS participants. 
% We believe this was primarily due to unfamiliarity of the field and we assume CS-participants only have skimmed through the physics article.

% \noindent
% \textbf{Pupil Dilation: }
While reading the physics article, we observed that CS participants have comparatively larger pupil dilation than Non-CS participants, potentially indicating domain unfamiliarity increasing their cognitive load. However, we did not notice much difference between participants while reading the computer science article.  

% 
% OLD Content
% 
% When reading the physics article, we observed that CS participants have comparatively larger pupil dilation compared to the Non-CS participants.
% This yields that the unfamiliarity of the field has increased their cognitive load while reading the article.
% On the contrary, we observed that the pupil dilation of both CS and Non-CS participants did not differ much when reading the 
% computer science article.

% \noindent
% \textbf{LHIPA: }
While reading the physics article, CS participants have experienced a higher cognitive load (lower average LHIPA) with the highest during the introduction. In contrast, Non-CS participants have experienced the highest in the methodology section with lesser cognitive load in other sections. In contrast, CS participants have undergone a higher cognitive load throughout the computer science article except abstract than Non-CS participants, despite our expectation of domain familiarity yielding a lower cognitive
load. 

% 
% OLD Content
% 
% When reading the physics article, CS participants have experienced a higher cognitive load (lower average LHIPA) during the introduction section, whereas Non-CS participants have experienced a higher cognitive load during the methodology section. 
% When reading the title and figures, both CS and Non-CS participants have experienced lower cognitive load (higher average LHIPA).

% When reading the computer science article, CS participants overall have experienced a comparatively high cognitive load throughout the paper, whereas Non-CS participants have experienced a comparatively low cognitive load throughout the paper.
% We expected Non-CS participants to experience comparatively higher cognitive load compared to CS participants while reading the computer science article.
% However, we observed that Non-CS participants have had the lower cognitive load when reading the title and figures of computer science article similar to when reading the physics article.

\section{Discussion and Conclusion}

We did not observe domain familiarity impacting cognitive load expressed through pupillometric characteristics based on the results. However, we observed higher fixations in abstract, introduction, and methodology irrespective of the domain familiarity. Moreover, we noticed a higher cognitive load during introduction and methodology irrespective of participant and article. 

% 
% OLD CONTENT
% 
% From the results, we observed that participants had comparatively larger pupil dilation when they read a paper which belongs to an unfamiliar field.
% We also observed comparatively higher fixation counts on sections including abstract, introduction, and methodology when participants read a paper which belongs to an unfamiliar field.
% OLD COMMENT : However, CS participants fixated much less on all of the sections of physics article compared to Non-CS participants. Our assumption is that CS-participants only have skimmed through the physics article. But this requires further investigation with a larger number of subjects.
% In addition, we observed a higher cognitive load from both types of participants while reading introduction and methodology sections in both articles.

Our preliminary results have multiple limitations. Firstly, we expect the observations to include potential biases resulting from the lack of diversity in our study sample due to the early stage of the study. We expect more generalizable gaze and pupillometric characteristics to emerge by diversifying the study participants. Further, we present only the most widely used measures in this study, while metrics beyond our study may uncover novel findings.

In the study, we performed manual annotations for mapping the gaze positions of the users to the sections of the paper, which is tedious and time-consuming for an experiment of large scale.
As presented in prior studies\cite{jayawardena2021automated-aoi, mahanama2021-reading-patterns-jcdl}, automated annotation approaches can help overcome the issue and form a novel research avenue. 
However, our experimental setup requires a clear point of view imagery and distinctive features in the digital documents to use those approaches effectively. 
Further, such an automated approach must be resilient to potential false positives in categorization. 

% We anticipate tailored literature summarization tools as a potential application of our findings. We can perform this by incorporating information-seeking behavior information into text summarization models, thus enforcing the model to focus more on the information sought by a researcher. However, the possible means of achieving it remains unexplored. 

% OLD CONTENT
% 
% Further, the preliminary results presented got multiple limitations. Firstly, we expect the observations presented in the paper to include potential biases resulting due to diversity limitations of our study sample due to the early stage of the study. Moreover, a more generalizable gaze and pupillometric characteristics might emerge by diversifying the study participants. Further, we present only the most widely used measures in this study, while metrics beyond our study may uncover novel findings. 

% OLD and REMOVED??
% Moreover, we anticipate the potential of using the information-seeking behavior characteristics for tailoring literature summarization for researchers. We can perform this by incorporating information-seeking behavior information we capture through eye-tracking into text summarization models, thus enforcing the model to focus more on the information sought by a researcher. However, the possible means of achieving it remains unexplored. 

\begin{acks}
This work is supported in part by the U.S. National Science Foundation grant CAREER IIS-2045523. Any opinions, findings and conclusion or recommendations expressed in this material are the author(s) and do not necessarily reflect those of the sponsors.
\end{acks}

\bibliographystyle{ACM-Reference-Format}
\bibliography{references}

% \appendix

% \section{Dataset}
\end{document}